# Intrinsic defects in photovoltaic perovskite variant $Cs_2SnI_6$

Zewen Xiao,[1,2] Yuanyuan Zhou,[3] Hideo Hosono,[1,2] Toshio Kamiya[1,2]

[1]Materials and Structures Laboratory, Tokyo Institute of Technology, Yokohama 226-8503, Japan

[2]Materials Research Center for Element Strategy, Tokyo Institute of Technology, Yokohama 226-8503, Japan

[3]School of Engineering, Brown University, Providence, RI 02912, United States

**ABSTRACT:** $Cs_2SnI_6$ is a variant of perovskite materials and is expected as a lead-free air-stable photovoltaic material. In this letter, we report intrinsic defects in $Cs_2SnI_6$ using first-principles density functional theory calculations. It is revealed that iodine vacancy and tin interstitial are the dominant defects that are responsible for the intrinsic *n*-type conduction in $Cs_2SnI_6$. Tin vacancy has a very high formation energy (>3.6 eV) due to the strong covalency in the Sn–I bonds and is hardly generated for *p*-type doping. All the dominant defects in $Cs_2SnI_6$ have deep transition levels in the band gap. It is suggested that the formation of the deep defects can be suppressed significantly by employing an I-rich synthesis condition, which is inevitable for photovoltaic and other semiconductor applications.

Metal halide perovskites have been introduced as the "game-changer" materials in the novel solid-state solar cells.[1–4] These perovskite compounds have the general chemical formula $ABX_3$ ($A$ = Cs, $CH_3NH_3$, or $CH_2NH{=}CH$; $B$ = Pb or Sn; $X$ = I, Br or Cl), where the $A$ cations sit in the cubic or pseudo-cubic network of corner-sharing $BX_6$ octahedra (see Fig. 1a). Among these materials, tin (Sn) based perovskites $ASnX_3$ have attracted particular interest due to their nontoxicity.[5–9] However, $ASnX_3$ are very sensitive to the ambient atmosphere (oxygen, moisture, etc.)[7–12] and exploration for air-stable alternatives has become an urgent issue. Promisingly, *Lee* et al.[13] recently reported that the Sn-based perovskite variant $Cs_2SnI_6$ exhibited high air-stability and their solar cells exhibit good efficiently as high as 8%. It is proposed that the superior stability of $Cs_2SnI_6$ to $CsSnI_3$ would be attributed to the +4 oxidation state of Sn in $Cs_2SnI_6$ based on a simple ionic model;[13] on the other hand, density functional / hybrid density functional theory (DFT/HDFT) calculations indicate that the real oxidation state of Sn in $Cs_2SnI_6$ is closer to +2, similar to $CsSnI_3$, which originates from the nominal formula $Cs^{+}{}_2Sn^{2+}I_6^{4-}$.[14] In this view, the improved stability of $Cs_2SnI_6$ is attributed to the strong covalency of the Sn–I bonds in the isolated $[SnI_6]^{2-}$ cluster.

It is known well that the typical $Sn^{2+}$-based compounds, including SnO,[15] SnS,[16] and $CsSnI_3$,[17] in general, intrinsically exhibit good *p*-type conductivity because $V_{Sn}$ in these compounds are easily formed and act as shallow acceptors to produce mobile holes. On the other hand, $Cs_2SnI_6$ have not shown *p*-type conductivity unlike the other $Sn^{2+}$-based compounds and have shown *n*-type conduction or insulating behavior depending on the synthetic routes. Lee *et al.*[13] reported the native *n*-type conduction with electron density ($n$) of ~$1\times10^{14}$ $cm^{-3}$ in polycrystalline $Cs_2SnI_6$ pellets annealed at 200 °C, while Zhang *et al.*[18] observed very high resistivity in room temperature-processed $Cs_2SnI_6$. Therefore, in order to clarify the intrinsic nature of $Cs_2SnI_6$ and its origin, a systematic theoretical study on the intrinsic defects is important, which will provide a guiding principle for tuning its properties for photovoltaic and other semiconductor applications.

Here, we studied the formation enthalpy ($\Delta H$) of intrinsic defects in $Cs_2SnI_6$ by DFT/HDFT calculations. It was clarified that I vacancy ($V_I$) and Sn interstitial ($Sn_i$) are mainly responsible for the intrinsic *n*-type conductivity in $Cs_2SnI_6$. $V_{Sn}$, which is usually a dominant *p*-type defect in $Sn^{2+}$-based compounds, is hardly formed in $Cs_2SnI_6$ because of its high $\Delta H$ which is caused by the strong covalency of the Sn–I bonds in the isolated $[SnI_6]^{2-}$ cluster. It was also clarified that intrinsic *p*-type conductivity is hardly be attained in pure $Cs_2SnI_6$ due to the absence of an effective acceptor with sufficiently low $\Delta H$ and a shallow transition level. The origin of deep transition levels of the dominant defects in $Cs_2SnI_6$ will be discussed.

$Cs_2SnI_6$ crystalizes into the anti-fluorite structure (space group $Fm\text{–}3m$, the lattice parameter $a$ = 11.65 Å[19]), which is composed of four $[SnI_6]^{2-}$ octahedra at the corners and the face centers and eight $Cs^+$ cations at the tetragonal interstitials, as shown in Fig. 1b. Alternatively, $Cs_2SnI_6$ can be regarded as a defective variant of the simple perovskite $CsSnI_3$ as seen by comparing Figs. 1b with 1a, which illustrates a 2×2×2 supercell of $CsSnI_3$ where the $[SnI_6]^{2-}$ octahedra connect to each other by sharing their corners. The $Cs_2SnI_6$ structure is obtained by removing a half of the Sn atoms at each center of the $[SnI_6]$ octahedron at intervals (i.e. those at the centers of orange octahedra in Fig. 1a) and relaxing the resulted structure, and thus the corner-shared $[SnI_6]^{2-}$ octahedra in Fig. 1a become isolated in $Cs_2SnI_6$ (Fig. 1b). It should be noted that the $[SnI_6]^{2-}$ octahedra shrink in the structural relaxation, leading to the smaller Sn−I bond length (2.85 Å)[19] in $Cs_2SnI_6$ than that in $CsSnI_3$ (3.11 Å).[9]

Defect calculations were performed for $Cs_2SnI_6$ in the framework of DFT/HDFT using the projector-augmented wave (PAW) method as implemented in the VASP code[20] (see the ESI† for details). Fig. 1c shows the calculated chemical potential ($\Delta\mu_{Sn}$, $\Delta\mu_I$) – phase map, in which the yellow region A–B–C–D indicates where $Cs_2SnI_6$ is stabilized against possible competitive phases including Cs, Sn, I, CsI, $SnI_2$, $SnI_4$, and $CsSnI_3$. The narrow shape indicates that the growth conditions should be carefully controlled to produce the single-phase $Cs_2SnI_6$, similar to the cases of $CsSnI_3$[17] and $MAPbI_3$.[21]

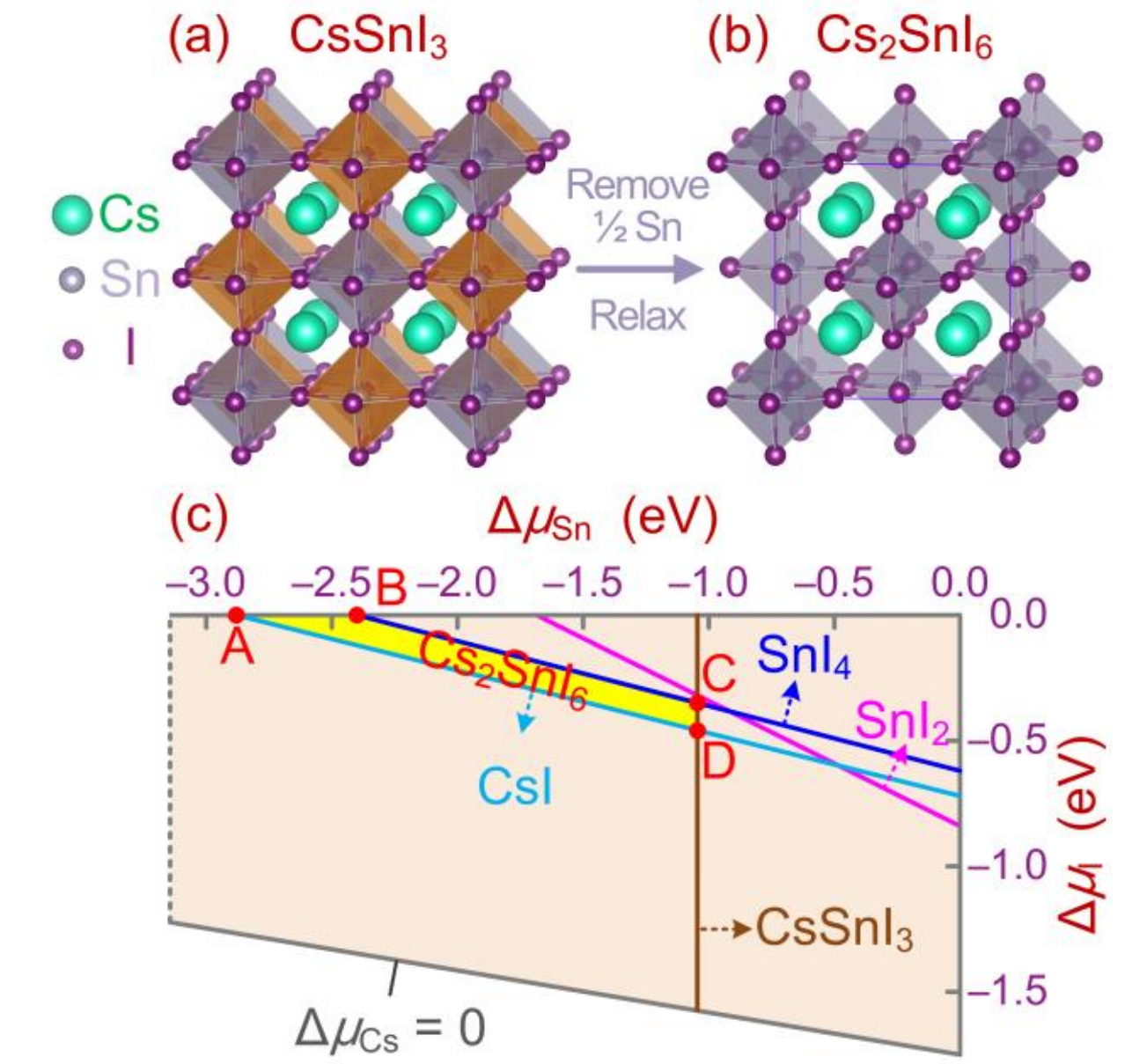

**Fig. 1.** (a) 2×2×2 supercell structure of $CsSnI_3$. (b) Crystal structure of $Cs_2SnI_6$ (cubic, space group *Fm*–3*m*, *a* = 11.65 Å) where Cs, Sn, and I atoms lie at the 8*c* (¼, ¼, ¼), 4*a* (0, 0, 0) and 24*e* (0.245, 0, 0) sites, respectively. (c) Chemical potential ($\Delta\mu_{Sn}$, $\Delta\mu_I$) – phase map. The yellow region A–B–C–D shows the region where $Cs_2SnI_6$ is stabilized against possible competitive phases including Cs, Sn, I, CsI, $SnI_2$, $SnI_4$, and $CsSnI_3$.

We have considered intrinsic point defects in $Cs_2SnI_6$ including three vacancies ($V_{Cs}$, $V_{Sn}$, $V_I$), three interstitials ($Cs_i$, $Sn_i$, $I_i$), two cation substitutions ($Cs_{Sn}$, $Sn_{Cs}$), and four antisites ($Cs_I$, $Sn_I$, $I_{Cs}$, $I_{Sn}$). Two representative chemical potential conditions in Fig. 1c are chosen for the following discussion; ($\Delta\mu_{Cs}$, $\Delta\mu_I$) at A (I-rich condition) and D (I-poor condition) points, where the calculated $\Delta H$ under are plotted as a function of the Fermi level ($E_F$) in Figs. 2a and 2b, respectively. The calculated transition levels $\varepsilon(q/q')$ are plotted relative to the conduction band minimum (CBM) and valence band maximum (VBM) in Fig. 3. Out of the twelve intrinsic defects, four ($V_I$, $Sn_i$, $Cs_I$ and $V_{Cs}$) have a sufficiently small $\Delta H$ (e.g., < 1.0 eV) to influence the electrical properties. Among them, $V_I$ has the lowest $\Delta H$ (≤0.74 eV and ≤0.28 eV under the I-rich and I-poor conditions, respectively) and acts as a deep donor with $\varepsilon(0/+1)$ = 0.74 eV above VBM (i.e., 0.52 eV below the conduction band edge, CBM), which is mainly responsible for the *n*-type condition of $Cs_2SnI_6$. Under the I-poor condition, $Sn_i$ has a small $\Delta H$ (≤1.18 eV) and a shallower transition $\varepsilon(0/+1)$ at 0.11 eV below the CBM. Therefore, $Sn_i$ as well as $V_I$ are also dominant donors for *n*-type conduction. $Cs_I$ is of a low $\Delta H$ but stabilized in neutral charge state at high $E_F$ (i.e. *n*-type conditions); therefore, it does not contribute to the *n*-type conduction. Only $Sn_I$ is a shallow donor in $Cs_2SnI_6$ with the $\varepsilon(0/+2)$ above the CBM, as seen in Figs. 2b and 3. However, because of its high $\Delta H$, $Sn_I$ has limited contribution to the *n*-type conductivity even under the I-poor condition. On the other hand, under the I-rich condition, $V_{Cs}$ has a low $\Delta H$ (≤1.37 eV) and can act as a deep acceptor, but at $E_F$ only above 0.51 eV from the VBM and the hole density produced is too low to compensate the electrons released by $V_I$.

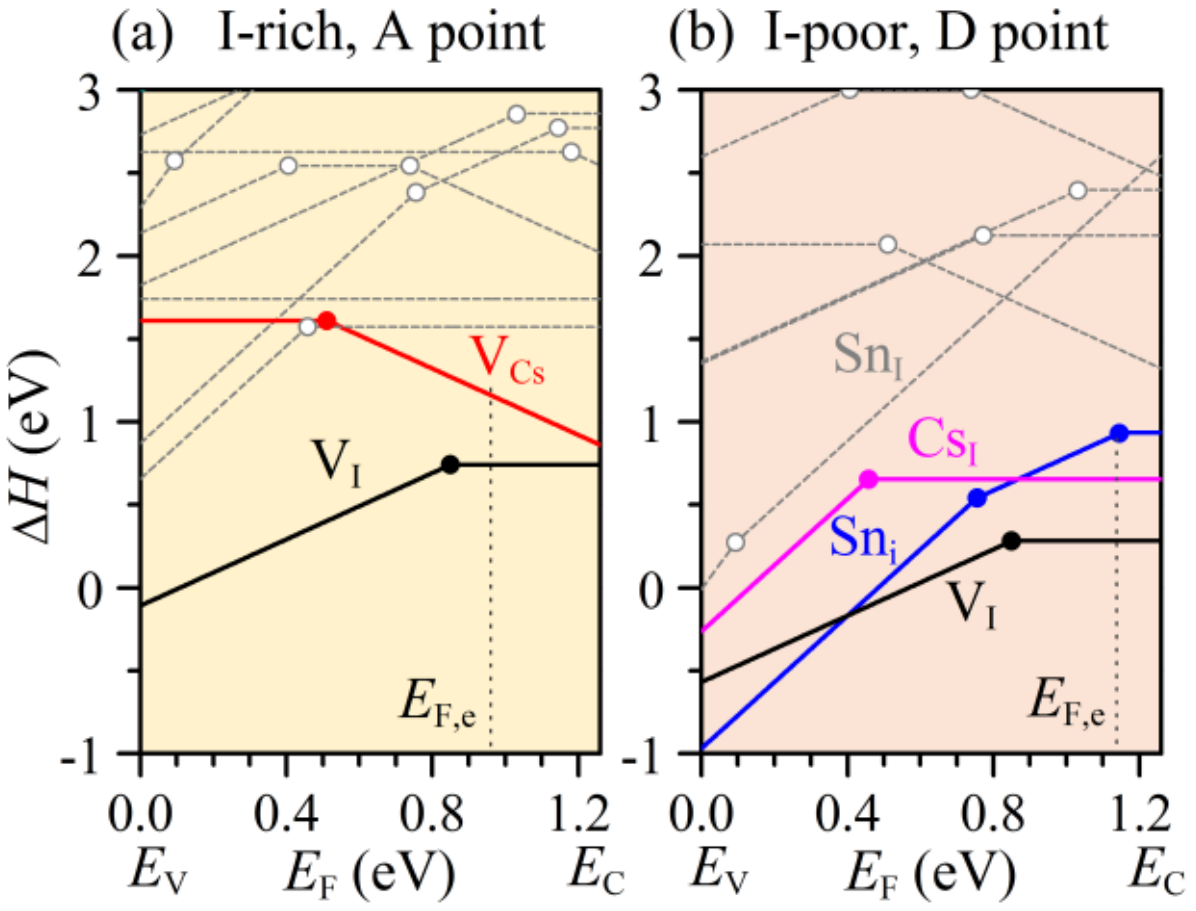

**Fig. 2.** Calculated $\Delta H$ of intrinsic defects in $Cs_2SnI_6$ as a function of $E_F$ at the chemical potential points in Fig. 1c, (a) A (I-rich) and (b) D (I-poor). Defects with much high $\Delta H$ values are shown by dashed lines.

The other intrinsic defects with prohibitively high $\Delta H$ are displayed by gray dashed lines in Fig. 2. It should be noted that $V_{Sn}$ in $Cs_2SnI_6$ has high $\Delta H$ values under all the chemical potentials and is as large as 3.63 eV even under the Sn-poor condition (A point), as seen in Table 1. This is strikingly different from the cases of the typical $Sn^{2+}$-based semiconductors such as SnO, SnS, and $CsSnI_3$, in which the $\Delta H$ values of $V_{Sn}$ (at the VBM) are usually very small under the Sn-poor condition (i.e., 1.3 eV at $E_V$ for SnO,[15] 0.8 eV for SnS,[16] and 0.3 eV for $CsSnI_3$[17]). As we previously reported, the Sn–I bonds in $Cs_2SnI_6$ exhibit a strong covalency (2.40 eV/bond) due to the much shortened Sn–I length in the isolated $[SnI_6]^{2-}$ octahedra.[14] Sn atoms are tightly encased in the $[I_6]$ octahedra and hardly removed by thermal fluctuation, which explains the unusually high $\Delta H$ of $V_{Sn}$ in $Cs_2SnI_6$. Due to the same reason, the substitutions of Sn by Cs and I (i.e., $Cs_{Sn}$ and $I_{Sn}$) have similarly high $\Delta H$ values. We have proposed that the +2 valence state of Sn in $Cs_2SnI_6$ is stabilized by the strong covalency of the Sn–I bonds.[14] Here, the high $\Delta H$ of the Sn-site-related defects further support the stabilized +2 valence state of Sn in $Cs_2SnI_6$.

Since it is found any intrinsic defect does not work as an effective *p*-type source from the I-rich to I-poor conditions, as shown above, it would be difficult to achieve intrinsic *p*-type conduction in pure $Cs_2SnI_6$. Instead, $Cs_2SnI_6$ intrinsically exhibits *n*-type conduction due to the easy formation of $V_I$ and $Sn_i$ donors. For quantitative analysis, the equilibrium $E_F$ ($E_{F,e}$) at room temperature were calculated by solving semiconductor statistic equations self-consistently so as to satisfy the charge neutrality condition (see the ESI† for details). Under the I-rich limit at the A point, $E_{F,e}$ is 0.96 eV above the VBM (see the vertical dotted line in Fig. 2a) and the corresponding electron density (*n*) is ~$10^{12}$ $cm^{-3}$, which is not easily determined by a Hall effect measurement. Under the I-poor condition at the D point, $E_{F,e}$ is 1.14 eV above the VBM (see the vertical dotted line in Fig. 2b) and *n* is ~$10^{16}$ $cm^{-3}$. These theoretical results seem to explain the experimental results reported to date; i.e., the electrical properties of $Cs_2SnI_6$ ranging from insulating[18] to low-density *n*-type conduction (*n* ~$1\times10^{14}$ $cm^{-3}$).[13]

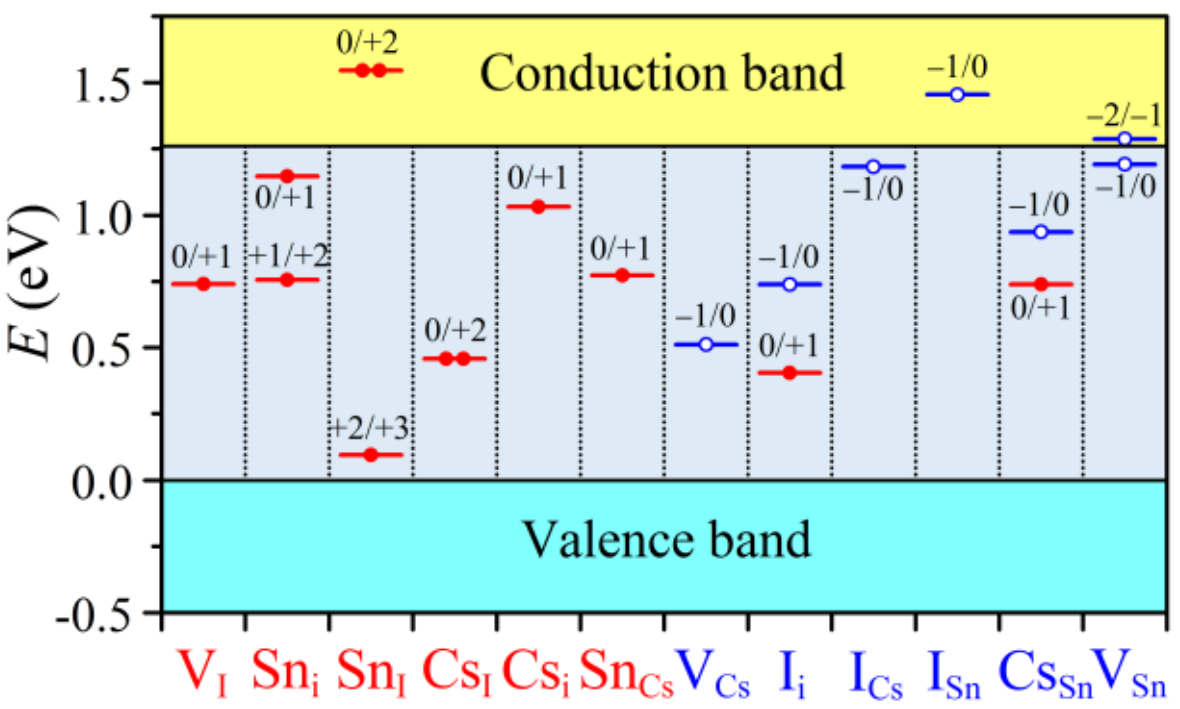

**Fig. 3.** Calculated transition energy levels $\varepsilon(q/q')$ for intrinsic defects in $Cs_2SnI_6$. Donor and acceptor defect levels are denoted by red and blue bars, respectively. The solid and open circles at the transition levels show the number of electrons and holes that can be released during the transition of the defect charge state.

**Table 1. Formation enthalpies (in eV) of twelve intrinsic defects in $Cs_2SnI_6$ at the chemical potential points A, B, C, and D shown in Figure 1.**

| | $V_{Cs}$ | $V_{Sn}$ | $V_I$ | $Cs_i$ | $Sn_i$ | $I_i$ | $Cs_{Sn}$ | $Sn_{Cs}$ | $Cs_I$ | $Sn_I$ | $I_{Cs}$ | $I_{Sn}$ |
|---|---|---|---|---|---|---|---|---|---|---|---|---|
| A | **1.61** | **3.63** | 0.74 | 2.86 | 2.77 | 2.54 | 3.74 | 3.50 | 1.57 | 5.47 | 2.62 | 1.74 |
| B | **1.37** | **4.11** | 0.74 | 3.10 | 2.28 | 2.54 | 4.46 | 2.78 | 1.81 | 4.98 | 2.38 | 2.23 |
| C | **1.71** | **5.47** | 0.40 | 2.76 | 0.93 | 2.88 | 5.48 | 1.76 | 1.14 | 3.29 | 3.06 | 3.92 |
| D | **2.07** | **5.47** | 0.28 | 2.40 | 0.93 | 3.00 | 5.12 | 2.12 | 0.65 | **3.17** | 3.54 | 4.04 |

The $\varepsilon(q/q')$ of the dominant defects are important in particular for solar cells because $\varepsilon(q/q')$ in the band gap indicates that they work as an electron trap, hole trap, and/or a recombination center, which deteriorate the device performances. The dominant defects $V_{Pb}$ in $MAPbI_3$[21] and $V_{Sn}$ in $CsSnI_3$[17] have transition levels deeper than their VBMs and are inert. Also in $CuInSe_2$, the dominant defect $V_{Cu}$ has an $\varepsilon(-1/0)$ at only 0.03 eV above the VBM.[22] The very shallow $V_{Cu}$ results in good *p*-type good conduction but does not deteriorate the photovoltaic performance. In contrast, in $Cs_2SnI_6$, all the dominant defects ($Cs_i$, $Sn_i$, $V_I$, $V_{Cs}$ as shown by the solid lines in Fig. 2) have $\varepsilon(q/q')$ in the band gap, as seen in Fig. 3, which would more or less hinder photovoltaic performances through short diffusion/drift length and fast recombination of photo-generated carriers, in particular when processed under the I-poor condition. These deep defects can, however, be significantly avoided by using I-rich conditions, which is located at the phase boundary between $Cs_2SnI_6$ and elemental I (e.g. the A point as shown in Fig. 2a), because their $\Delta H$ are considerably increased as seen in Table 1.

The defect properties as well as electronic structure of a material can qualitatively understood well based on the molecular orbital theory.[21,23,24] Here we provides a simplified interpretation for the origin of the deep nature of the intrinsic defects in $Cs_2SnI_6$ in comparison with representative $Sn^{2+}$-based compounds such as SnO and $CsSnI_3$, by focusing on cation and anion vacancies, which generally have low $\Delta H$ and play important role on electrical properties of these compounds. Figure 4a shows the calculated densities of states (DOSs) and projected DOSs (PDOSs) for SnO, $CsSnI_3$ and $Cs_2SnI_6$, where the energy is aligned by Sn 4*d* and Cs 5*s* so as to compare the energy levels. Figures 4b–4d show the schematic energy diagrams that are derived From Fig. 4a. As depicted in Figure 4b, the VBM of SnO consists of the antibonding states of Sn 5*s* and O 2*p* orbitals (region I in the top panel of Fig. 4a). For a cation vacancy $V_M$, it is known that the $V_M$ transition energy in a simple oxide is ~1 eV above VBM e.g. in ZnO.[25] Also for the SnO case, the $V_{Sn}$ level is above the O 2*p* band (region II of the top panel of Fig. 4a) similar to the $V_M$ in the simple oxides. As the VBM made by Sn 5s level is similarly high, and consequently the $V_{Sn}$ level in SnO forms the shallow acceptor level. On the other hand, the CBM consists of antibonding Sn 5*p*–O 2*p* states and is close to the Sn 5*p* level due to the ionic character of SnO. For an oxygen vacancy $V_O$, the defect state is composed of the Sn non-bonding state, which drops slightly from the CBM,[23] resulting in the shallow nature of $V_O$. Fig. 2c schematically illustrates the formation of the CBM, VBM, $V_{Sn}$, and $V_I$ in $CsSnI_3$. Since the Cs cation does not contribute to the electronic structure around the band gap, the electronic structure of $CsSnI_3$ is similar to that of SnO if O is replaced with I. As a result, $V_{Sn}$ and $V_I$ in $CsSnI_3$ form a shallow acceptor and a shallow donor, respectively, similar to those in SnO.[17]

In contrast, $Cs_2SnI_6$ has a strikingly different electronic structure (see the bottom panel of Fig. 4a and Ref. 14 for details), which is schematically illustrated in Fig. 4d. Because the removal of the half of Sn atoms and the subsequent isolation of the $[SnI_6]^{4-}$ octahedra, the energy splitting between the bonding and the antibonding states of Sn 5*s*–I 5*p* (see regions IV and I, respectively, in the bottom panel of Fig. 4a) become very large, and the Sn 5*s*–I 5*p* bonding energies are deepened; on the other hand, the Sn 5*s*–I 5*p* antibonding energies are pushed up and become unoccupied states, forming the CBM of $Cs_2SnI_6$. The VBM is composed of the I 5*p*–I 5*p* antibonding states (region II in the bottom panel of Fig. 4a). It should be noted that the Sn 5*s* level in $Cs_2SnI_6$ is deep similar to SnO and $CsSnI_3$ and almost fully occupied, indicating the +2 oxidation state of Sn. It is seen that the $V_{Sn}$ level in $Cs_2SnI_6$ is similar to those in SnO and $CsSnI_3$, while the VBM level in $Cs_2SnI_6$ is far below that in $CsSnI_3$; consequently, the $V_{Sn}$ in $Cs_2SnI_6$ is a deep acceptor even close to the CBM, which explains why $Cs_2SnI_6$ is not an intrinsic *p*-type semiconductor.

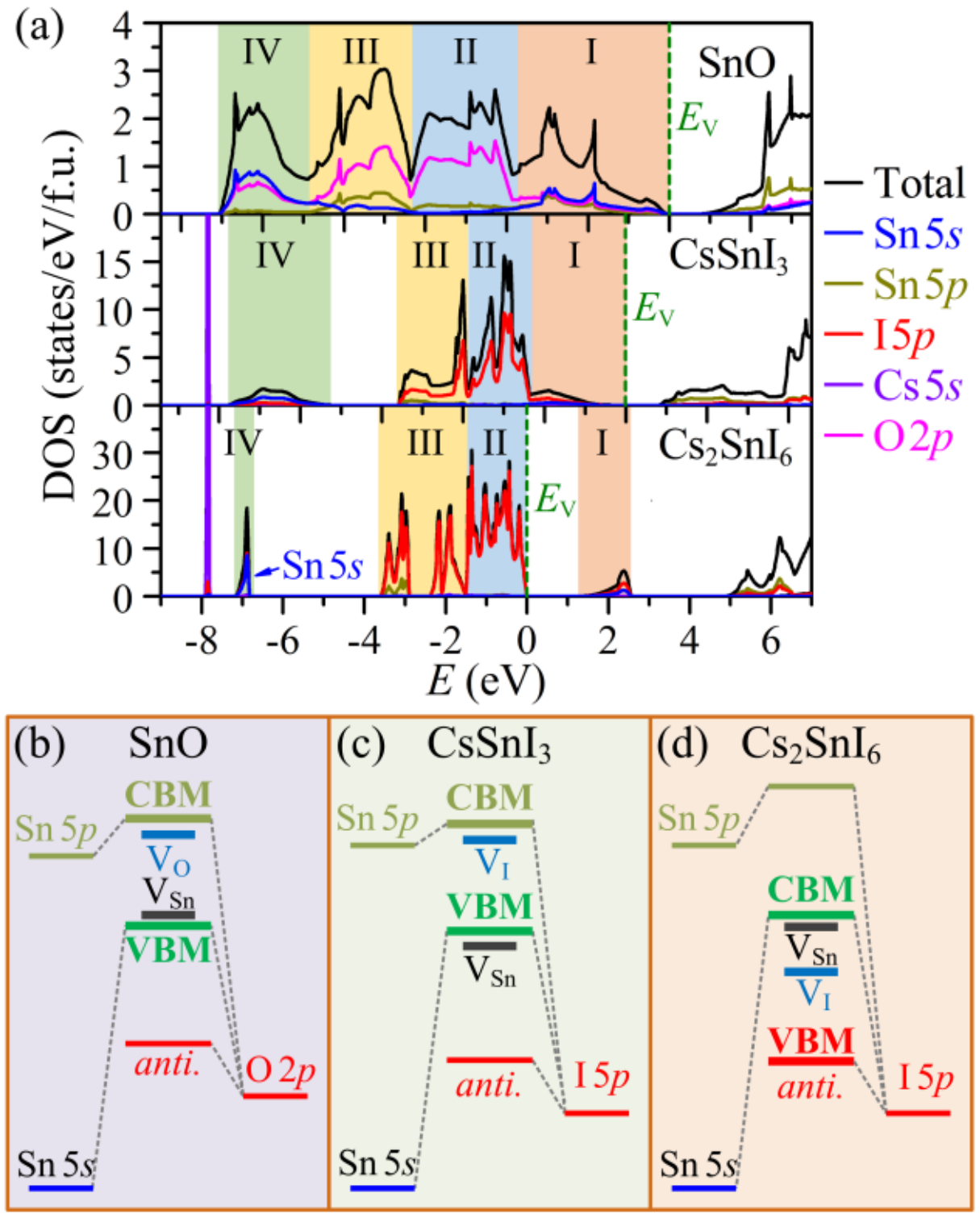

**Fig. 4.** (a) Total and projected densities of states (DOSs/PDOSs) of $Sn^{2+}O^{2-}$ (top panel), $Cs^{+}Sn^{2+}I^{-}_3$ (middle panel), and $Cs^{+}_2Sn^{2+}I_6^{4-}$ (bottom panel). Simplified energy diagrams depicting the formation of VBM, CBM, and donor-/acceptor-like defects in (b) $Sn^{2+}O^{2-}$, (c) $Cs^{+}Sn^{2+}I^{-}_3$, and (d) $Cs^{+}_2Sn^{2+}I_6^{4-}$.

In summary, we have investigated the formation energies of intrinsic defects in the $Cs_2SnI_6$ perovskite variant. The dominant defects are donor $V_I$ and $Sn_i$, giving the calculated electron densities of $10^{12}$–$10^{16}$ cm$^{-3}$. Unlike other $Sn^{2+}$-based *p*-type semiconductor such as SnO, SnS and $CsSnI_3$, the $V_{Sn}$ in $Cs_2SnI_6$ has a very high $\Delta H > 3.6$ eV and are hardly formed, primarily because of the strong Sn–I covalent bonds in the functional group-like $[SnI_6]^{2-}$ isolated cluster. The energy levels of the dominant defects are deep in the band gap and would work as recombination centers in photovoltaic applications. On the other hand, their formations would be practically suppressed by employing an I-rich synthesis condition where their densities are negligibly small due to the high $\Delta H$. These results explain the reported experimental results and provide a clue to better understanding the unusual defect physics in *p*-block metal-based compounds.

This work was conducted under Tokodai Institute for Element Strategy (TIES) funded by MEXT Elements Strategy Initiative to Form Core Research Center. Y.Z. thanks Prof. Nitin P. Padture and U.S. National Science Foundation (DMR-1305913) for the support.